\documentclass[twocolumn,showpacs]{revtex4}
\usepackage{graphicx}
\usepackage{bm}
\usepackage{color}
\usepackage{amsmath}
\usepackage{natbib}

\begin{document}

\title{Exact solution for large amplitude circularly polarized electromagnetic waves in incompressible spin quantum Hall magnetohydrodynamics}

\author{Felipe A. Asenjo}
	\email{fasenjo@levlan.ciencias.uchile.cl}
\affiliation{Departamento de F\'\i sica, Facultad de Ciencias, Universidad de Chile, Casilla 653, Santiago, Chile.}
\affiliation{Departamento de Ciencias, Facultad de Artes Liberales, Universidad Adolfo Ib\'a\~nez, Diagonal Las Torres 2640, Pe\~nalol\'en, Santiago, Chile.}

\date{\today}

\begin{abstract}
It is shown that incompressible spin quantum Hall magnetohydrodynamics allows an exact solution for the propagation of a circularly polarized electromagnetic wave. The solution is obtained assuming a condition between the fluid velocity and the magnetic field which eliminates the nonlinear terms in Maxwell equations. As a result of the coupling with spin, the propagation mode depends on the amplitude of the magnetic field. From the full solution, the limits of small and large wavenumber are studied obtaining linear and a nonlinear spin-modified modes. 
\end{abstract}

\maketitle

\section{Introduction}

Quantum plasmas have produced huge interest in recent years because they can be important in high-energy scenarios. For a plasma with number density $n$, the quantum effects are relevant when the thermal de Broglie wavelength $\lambda_B$ of the plasma constituents is similar to or larger than the average interparticle distance $n^{-1/3}$, i.e. when $n\lambda_B^3\geq1$ \cite{manfredi}. 

When the spin of the electrons is considered, a fluid plasma theory can be constructed starting from the Pauli Hamiltonian \cite{brodin2}. Using this fluid description, a magnetohydrodynamics (MHD) theory for spin quantum plasmas can be obtained \cite{brodin}. The main quantum corrections of these theories are the so called Bohm potential which is due to density fluctuations, the contribution of the spin of electrons, and the thermal-quantum terms couplings. When the spin is neglected, only quantum contributions associated to density fluctuations, as Bohm potential, corrects the quantum plasma equations \cite{haas}.

Many studies have been done to find the quantum corrections to different linear and nonlinear propagation  modes  at first order perturbations in quantum plasmas and spin quantum plasmas \cite{shuk,ren,shuk3,saleem,haas2,ali,shuk2,misra}.
However, what are the spin quantum corrections to exact propagation modes? The aim of this work is to answer that question. Exact solutions for large amplitude electromagnetic (EM) waves can be found in classical and relativistic fluid plasma theories. For instance, a particular case of exact solution of plasma equations are the propagation of circularly polarized EM waves in cold and hot magnetized relativistic plasmas \cite{barnes,gomberoff,asenjo}. 

The difficulty of finding exact solutions grows with the increasing complexity introduced by the different effects considered.
Recently, Mahajan and Krishan \cite{mahaj} showed that an exact solution for propagation of a circularly polarized EM wave can be obtained for an ideal incompressible Hall MHD (HMHD) theory \cite{huba}. The solution is constructed assuming a Beltrami relation between magnetic field and the fluid velocity. This is the first nonlinear large amplitude solution that is found in this regime. 

In this work, we derive the exact solution for the propagation mode of a circularly polarized EM wave in a spin quantum plasma in the MHD regime considering the Hall current density effect in the Ohm law. We use the MHD formalism developed in Ref.~\cite{brodin} for spin quantum plasmas.  The quantum effects on the exact propagation mode of circularly polarized EM waves and on their nonlinear effects have been studied previously in magnetized quantum plasmas \cite{stenflo} and in magnetized spin quantum plasmas \cite{marklund}, both without the Hall current density effect.

\section{Exact non-linear solution}

\subsection{Spin Hall MHD}

Fluid formalisms for spin quantum plasmas are constructed using the Pauli Hamiltonian \cite{brodin, brodin2}. From this Hamiltonian, it is possible to obtain fluid equations with quantum corrections for the density and velocity of the plasma. Besides, the spin has a dynamical equation which is coupled to the velocity. 
The full description for the spin quantum HMHD (QHMHD) \cite{brodin} can be obtained from an electron-ion plasma where the quantum corrections are due to the electron. 
Neglecting the quantum-pressure couplings, the nonlinear inviscid QHMHD theory with spin contributions is given by the continuity equation
\begin{equation}
 \frac{\partial\rho}{\partial t}+\nabla\cdot(\rho{\bf v})=0\, ,
\label{contH}
\end{equation}
where $\rho$ is the one-fluid mass density and ${\bf v}$ is the one-fluid velocity; the momentum equation
\begin{equation}
 \rho\left(\frac{\partial}{\partial t}+{\bf v}\cdot\nabla\right){\bf v}=\frac{{\bf J}}{c}\times{\bf B}+\nabla p+{\bf F}_Q\, ,
\label{momeH}\end{equation}
where ${\bf J}$ is the current density, $c$ the speed of light in vacuum, ${\bf B}$ is the magnetic field, $p$ is the pressure, and ${\bf F}_Q$ are the quantum force density contributions; and the Maxwell equation 
\begin{equation}
 \frac{\partial{\bf B}}{\partial t}=\nabla\times\left({\bf v}\times{\bf B}-\frac{1}{en_e c}{\bf J}\times{\bf B}\right)\, ,
\label{MaxwH}\end{equation}
where $e$ is the electric charge, $n_e$ is the density of the electron fluid and we have neglected the resistivity and the pressure contribution of electrons. The last term on the right-hand side of Eq.~\eqref{MaxwH} is the Hall current term which comes from the Ohm's law. The Hall effect appears due to the velocity difference between electrons and ions when kinetic effects are not
considered, and it is relevant at length scales shorter than the ion
inertial length and time scales of the order, or shorter, than the ion cyclotron period \cite{galtier}.

In the QHMHD theory, the spin has a dynamical equation \cite{brodin}. Here, we consider a static solution such that the spin of the electrons is ${\bf s}=-\left(\hbar/2\right)\eta(\alpha){\bf B}/{B}$ (with $B=|{\bf B}|$), which is antiparallel to the magnetic field and minimizes the magnetic moment energy. The function $\eta(x)\equiv\tanh(x)$  is the Brillouin function due to the magnetization of a spin distribution in thermodynamic equilibrium with $\alpha= \mu_B B/(k_B T_e)$. The magnetic moment is $\mu_B=e\hbar/(2 m_e c)$ with the electron mass $m_e$, the temperature of the electron fluid $T_e$ and Boltzmann constant $k_B$. The funtion $\eta(\alpha)$ appears as the solution of the spin evolution equation for spin quantum plasmas where the spin inertia and the spin-thermal coupling terms are neglected \cite{brodin}. In the limit of cold plasma, $\eta(\alpha)=1$.

The current density of this system is given by
\begin{equation}
 {\bf J}=\frac{c}{4\pi}\nabla\times{\bf B}-c\nabla\times{\bf M}\, ,
\end{equation}
and ${\bf M}$ is the magnetization density produced in the medium by the spin of electrons
\begin{equation}
 {\bf M}=n_e \mu_B  \eta(\alpha) \frac{{\bf B}}{B}\, .
\label{magH}
\end{equation}

Finally, the quantum force of Eq. \eqref{momeH} is given by
\begin{equation}
 {\bf F}_Q=\frac{\hbar^2 n_e}{2 m}\nabla\left(\frac{\nabla^2\sqrt n_e}{\sqrt n_e}\right)+n_e\eta(\alpha)\mu_B \nabla B\, .
\label{qforce}
\end{equation}
The first term is the force due to the Bohm potential and the second term is the quantum force due to the spin.

The above formalism have been used for the study of the one-dimension nonlinear evolution of Alfv\'en waves in HMHD with spin quantum effects \cite{gosh}, showing that the spin can produce a considerable modification in the dynamics of the DNLS equation.

\subsection{Large amplitude circularly polarized wave}

Consider now a large amplitude circularly polarized magnetic field in a magnetized plasmas. 

Here, the circularly polarized magnetic field is represented by ${\bf B}_\perp(z,t)=B_\perp [\cos(ikz-i\omega t)\hat x+\sin(ikz-i\omega t)\hat y]$, with constant  amplitude $B_\perp$. The total magnetic field is ${\bf B}(z,t)={\bf B}_\perp(z,t)+B_0\hat z$, where $B_0$ is the  constant amplitude of the background magnetic field. 
The circularly polarized magnetic field propagates along the direction of the background magnetic field such that ${\bf B}_\perp\cdot{\bf k}=0$, with ${\bf k}=k\hat z$.

The circularly polarized wave induces a circularly polarized transverse velocity in the fluid ${\bf v}_\perp=v_\perp [\cos(ikz-i\omega t)\hat x+v_y \sin(ikz-i\omega t)\hat y]$, with constant  amplitude $v_\perp$. Due to $\nabla\cdot{\bf v}_\perp=0$, the fluid is incompressible, and from continuity equation \eqref{contH} we obtain that the density $\rho$ is constant.
If we assume that the ion and electron density $n_e$ are constants, then the Bohm potential in \eqref{qforce} vanish, and the quantum force becomes longitudinal depending on the magnetic field only.
Thus, we can write the momentum equation \eqref{momeH} for the transversal part of the circularly polarized velocity. Using the space time dependence of the circularly polarized quantities, we find
\begin{equation}
 -i\omega\rho{\bf v}_\perp=ik\frac{B_0}{4\pi}{\bf B}_\perp-ikB_0 {\bf M}_\perp\, .
\label{mom2H}
\end{equation}
Here, ${\bf M}_\perp$ is the transverse part of the magnetization density \eqref{magH}
\begin{equation}
 {\bf M}_\perp=n_e \mu_B  \eta(\alpha) \frac{{\bf B}_\perp}{B}\, ,
\label{magHt}
\end{equation}
where now $\alpha= \mu_B B/(k_B T_e)$, and $B\equiv|{\bf B}|=\left(B_\perp^2+B_0^2\right)^{1/2}$ is constant. 
We can notice that the quantum force \eqref{qforce} does not contribute to the momentum equation. Instead, the effect of the spin is introduced through the magnetization current density. 

The momentum equation \eqref{mom2H} is very similar to those obtained for Alfv\'en propagation. To find an exact solution for the propagation of the large amplitude circularly polarized wave in a plasma with Hall effect, we use the ansatz introduced in Ref.~\cite{mahaj} to include the Hall effect and the spin contributions in the dispersion relation. We assume that the general dispersion relation is given by
\begin{equation}
 \omega = \Lambda c_A k\, ,
\label{dispH}
\end{equation}
where $c_A=B_0/\sqrt{4\pi\rho}$ is the Alfv\'en velocity, and $\Lambda\equiv\Lambda(k)$ is an adimensional function which contains the Hall and spin effects and depends on the wavenumber $k$. Using the dispersion relation \eqref{dispH} in Eq.~\eqref{mom2H}, we can obtain a closed expression between the circularly polarized velocity and the magnetic field
 \begin{equation}
 {\bf v}_\perp=\frac{-{\bf B}_\perp}{\Lambda\sqrt{4\pi\rho}}\left(1-\frac{4\pi\mu_B n_e \eta(\alpha)}{B} \right)\, .
\label{mom3H}
\end{equation}

The problem is reduced to find the value of the $\Lambda$ function. This is achieved using the Maxwell equation \eqref{MaxwH}, written for the circularly polarized quantities
\begin{widetext}
\begin{equation}
 \frac{\partial{\bf B}_\perp}{\partial t}=\nabla\times\left[\left({\bf v}_\perp-\frac{1}{4\pi e n_e}\left(1-\frac{4\pi\mu_B n_e\eta(\alpha)}{B}\right)\nabla\times{\bf B}_\perp\right)\times\left({\bf B}_\perp+B_0\hat z\right)\right]\, .
\label{MaxwH2}\end{equation}

The solution of \eqref{MaxwH2} and \eqref{mom3H} is complicated in general. However, this nonlinear problem can be converted in linear one imposing a condition between the velocity and the magnetic field. Following Ref.~\cite{mahaj}, a solution of the Maxwell equation \eqref{MaxwH2} is obtained when the velocity and the magnetic field satisfy the relation
\begin{equation}
 {\bf v}_\perp-\frac{1}{4\pi e n_e}\left(1-\frac{4\pi\mu_B n_e\eta(\alpha)}{B}\right)\nabla\times{\bf B}_\perp=\frac{-\Lambda}{\sqrt{4\pi\rho}}{\bf B}_\perp\, .
\label{conditionH}\end{equation}

Using this condition and the dispersion relation \eqref{dispH}, we remove the non-linear terms from Maxwell equation \eqref{MaxwH2}, which is now solved identically. This is because of circularly polarized magnetic field \eqref{mom3H} and Eq.~\eqref{conditionH} produce a Beltrami equation  for the magnetic field \cite{yoshida}. This is the necessary condition to obtain a dispersion relation with the form of Eq.~\eqref{dispH}.
Using the relation \eqref{mom3H} in Eq.~\eqref{conditionH}, we obtain the Beltrami equation for the magnetic field
\begin{equation}
 \frac{1}{4\pi e n_e}\left(1-\frac{4\pi\mu_B n_e\eta(\alpha)}{B}\right)\nabla\times{\bf B}_\perp=\frac{{\bf B}_\perp}{\sqrt{4\pi\rho}}\left(\Lambda-\frac{1}{\Lambda}+\frac{4\pi\mu_B n_e\eta(\alpha)}{\Lambda B}\right)\, ,
\end{equation}
which is a equation for the components of the circularly polarized magnetic field. Owing to $\nabla\times{\bf B}_\perp=ik\hat z\times{\bf B}_\perp$, this equation can be solved for $\Lambda$ to find two solutions
\begin{equation}
 \Lambda_{\pm}=\frac{1}{2}\left(1-\frac{4\pi \mu_B n_e\eta(\alpha)}{B}\right)\left[-\frac{k}{e n_e}\sqrt{\frac{\rho}{4\pi}}\pm\left(\frac{\rho k^2}{4\pi e^2 n_e^2}+4\left(1-\frac{4\pi\mu_B n_e\eta(\alpha)}{B}\right)^{-1} \right)^{1/2}\right]\, .
\label{lambda}
\end{equation}
\end{widetext}

The exact solution for the propagation of the circularly polarized EM wave in the QHMHD is given by the dispersion relation \eqref{dispH} with $\Lambda$ of Eq.~\eqref{lambda}. The contribution of the spin in the $\Lambda$ term are proportional to $\mu_B$, while the effect of the Hall current is proportional to $k$. Notice that owing to the spin, the exact solution $\Lambda_\pm$ depends on the amplitude $B$ of the total magnetic field.

\subsection{Analysis of the dispersion relation}

The nonlinear dispersion relation \eqref{dispH} can be studied in general for the wide-range of the wavenumbers. However, it is useful to studied this full solution in some extreme cases to see the contribution of the Hall current density. 

First, we will focus our study to the different known limit cases. If both spin and Hall effects are neglected, then $\Lambda_\pm=\pm 1$, and, therefore, we recover the simple Alfv\'en mode $\omega=\pm c_A k$. This is in agreement with that Alfv\'en modes are an exact solution of the MHD equations for a large amplitude circularly polarized EM wave. This occurs because the velocity and the magnetic field are parallel or anti-parallel, and therefore, the nonlinear terms vanish.

On the other hand, if only the spin is neglected ($\mu_B=0$), then we obtain
\begin{equation}
 \Lambda_{\pm}=-\frac{k}{2 e n_e}\sqrt{\frac{\rho}{4\pi}}\pm\frac{1}{2}\left[\frac{\rho k^2}{4\pi e^2 n_e^2}+4 \right]^{1/2}\, ,
\end{equation}
recovering the result of Ref. \cite{mahaj} for a circularly polarized EM wave in classical HMHD, where there are no dependence on the amplitude of the magnetic fields.

Now, we can study the contribution of the spin to the mode. 
The full solution \eqref{lambda} is for any value of $k$. The nonlinear dependence of $k$ in the dispersion relation and in the relation between velocities and magnetic fields is the most remarkable characteristic of this exact wave in the Hall regime. 
As in the classical case, we can examine the spin contribution to different opposite extreme cases for wavenumbers.
We consider first Eq.~\eqref{lambda} in the long-wavelenght limit $k\ll e n_e\sqrt{4\pi/\rho}$, which correspond to the MHD regime. To take this approximation is the same that to neglect the Hall current effect. We can find the exact solution for the dispersion relation $\omega=\pm {\tilde c}_A k$, where ${\tilde c}_A$ is a spin-modified Alfv\'en velocity given by
\begin{equation}
{\tilde c}_A=c_A\left(1-\frac{4\pi\mu_B n_e\eta(\alpha)}{B}\right)^{1/2}\, . 
\end{equation}

This is the spin quantum corrected mode of the shear wave in MHD. This propagation mode contains the spin correction to the classical Alfv\'en mode. If $\mu_B B\ll k_B T_e$, then $\eta(\alpha)\approx\alpha$, and ${\tilde c}_A=c_A\left[1-\hbar^2 \omega_p^2/(4m_e c^2 k_B T_e)\right]^{1/2}$, where $\omega_p=(4\pi e^2 n_e/m_e)^{1/2}$ is the plasma frequency of the electron fluid. 

In the opposite limit, the Hall-dominated regime is obtained when $k\gg e n_e\sqrt{4\pi/\rho}$ is considered in \eqref{lambda}. In this limit, it is possible to find the two modes of propagation
\begin{equation}
 \omega_+=\sqrt{\frac{4\pi}{\rho}}{e n_e c_A}\, ,
\label{kgrande1}
\end{equation}
and
\begin{equation}
 \omega_-=-\sqrt{\frac{\rho}{4\pi}}\frac{{{\tilde c}_A}^2}{e n_e c_A}k^2\, .
\label{kgrande2}
\end{equation}

Dispersion relation \eqref{kgrande1} represents plasma oscillations which do not propagate. This mode does not have any spin correction and coincides with the classic result of Ref. \cite{mahaj}. The relation  \eqref{mom3H} between the velocity of the fluid and the magnetic field becomes in ${\bf v}_\perp\propto -{\bf B}_\perp k {\tilde c}_A^2/c_A^2$, showing that they are aligned and depend on the spin contribution through the spin-modified Alfv\'en velocity. As in the classical treatment, this mode can be recognized with the propagation mode for the magnetosonic-cyclotron branch in the classical HMHD \cite{mahaj}. 

The mode \eqref{kgrande2} depends on the spin of the electrons through ${\tilde c}_A^2/c_A$. This is an exact nonlinear propagating mode which can be identified with the shear-whistler mode in classical HMHD \cite{mahaj}. The relation between its velocity and the magnetic field is ${\bf v}_{\perp}\propto{\bf B}_\perp/k$. They are still aligned and do not depend on the spin. Besides, for large values of $k$,  the velocity of this mode decrease.

\section{Discussion}

We have found the exact solution for a circularly polarized EM wave in an incompressible HMHD with spin quantum corrections. This result generalize previous ones for classical HMHD \cite{mahaj}, and at the knowledge of the author is the first exact solution found for QHMHD. The main difference with respect to classical result is that the solution for the wave propagation [Eqs.~\eqref{dispH} and \eqref{lambda}] depends on the amplitude of the magnetic field $B$. This is due to the form of the coupling between the spin and magnetic fields. 

Usually, the spin correction is important in very dense plasmas with strong background magnetic fields \cite{brodin2,gosh,marklund}. Under that conditions, relativistic effects might be important. For instance, with a density of $n_e\sim 10^{24}\mbox{cm}^{-3}$, an electron temperature given by $T_e=\hbar e B/(k_B m_e c)$ \cite{marklund}, and magnetic fields of $B_0\sim 10^{13}\mbox{gauss}$ and $B_\perp\sim 1~\mbox{gauss}$, we have that $1-{\tilde c}^2_A/c^2_A\simeq5\times 10^{-9}$. This correction is very small in general and it can be difficult to detect. 
However, it has been shown theoretically that spin can be important for nonrelativistic modes of propagation when $n_e\gg 10^{26}\mbox{cm}^{-3}$ and $B<4.4138\times 10^{13}\mbox{gauss}$ \cite{marklund}. If we choose a density of $n_e\sim 10^{30}\mbox{cm}^{-3}$ and a magnetic fields of $B_0\sim 10^{11}\mbox{gauss}$, at the same temperature than previous example, then ${\tilde c}^2_A/c^2_A\simeq 0.46$. This is very relevant for the group velocity  of this new waves. For instance, it represents an a $32\%$ of decreasing in the Alfv\'en velocity of the long-wavelenght mode (with $k\ll n_e\sqrt{4\pi/\rho}$), and a $54\%$ of decreasing for the $\omega_-$ mode (with $k\gg n_e\sqrt{4\pi/\rho}$), both with respect to their classical values.

As a final remark, it is important to say that exact solutions of classical HMHD can play a relevant role in turbulence studies \cite{galtier}. To know the correct relation between velocity and magnetic field is very important for the nonlinear dynamics of the turbulent plasma. Thus, exact solutions \eqref{lambda}, \eqref{kgrande1} and \eqref{kgrande2} in QHMHD could bring new insights in the nonlinear dynamics of plasmas with spin quantum contributions \cite{gosh}.

\acknowledgments

The author thanks to P. Zapata and R. Asenjo for their initial support.

\end{document}